\documentclass[twocolumn,preprintnumbers,amsmath,amssymb,superscriptaddress]{revtex4-1}
\usepackage{epsfig}
\usepackage{graphicx}
\usepackage{dcolumn}
\usepackage{bm}
\usepackage{float}
\usepackage{color}

\begin{document}

\selectfont

\title{Optimal box-covering algorithm for fractal dimension of complex networks}

\author{Christian M. Schneider}
\email{schnechr@mit.edu}
\affiliation{Computational Physics, IfB, ETH Zurich, Schafmattstrasse 6, 8093 Zurich, Switzerland}
\affiliation{Department of Civil and Environmental Engineering, MIT, 77 Massachusetts Avenue, Cambridge, MA 02139, USA}
\author{Tobias A. Kesselring}
\affiliation{Computational Physics, IfB, ETH Zurich, Schafmattstrasse 6, 8093 Zurich, Switzerland}
\author{Jos\'e S. Andrade Jr.}
\affiliation{Departamento de F\'{\i}sica, Universidade Federal do Cear\'a, 60451-970 Fortaleza, Cear\'a, Brazil}
\author{Hans J. Herrmann}
\affiliation{Computational Physics, IfB, ETH Zurich, Schafmattstrasse 6, 8093 Zurich, Switzerland}
\affiliation{Departamento de F\'{\i}sica, Universidade Federal do Cear\'a, 60451-970 Fortaleza, Cear\'a, Brazil}

\date{\today}
 
\begin{abstract}
The self-similarity of complex networks is typically investigated through computational algorithms the primary task of which is to cover the structure with a minimal number of boxes. Here we introduce a box-covering algorithm that not only outperforms previous ones, but also finds optimal solutions. For the two benchmark cases tested, namely, the {\it E. Coli} and the WWW networks, our results show that the improvement can be rather substantial, reaching up to $15\%$ in the case of the WWW network.

\end{abstract}

%%%%PACS e Keywords
 

\maketitle
\section{Introduction}
The topological and dynamical aspects of complex networks have been the focus of intensive research during the last years \cite{Watts1998,Albert1999,Barthelemy1999,Lloyd01,Cohen2003,Barthelemy04,Gonzalez2006,Gallos2007,Moreira2009,Hooyberghs2010,Li2010,Herrmann2011,Schneider2011b,Schneider2011a,Vespignani2012}. An open and unsolved problem in network and computer science is the following question: how to cover a network with the fewest possible number of boxes of a given size \cite{Peitgen1993,Feder1988,Bunde1995,Jensen1995,Cormen2001,song2006a}? In a complex network, a box size can be defined in terms of the chemical distance, $l_B$, which corresponds to the number of edges on the shortest path between two nodes. This means that every node is less than $l_B$ edges away from another node in the same box. Here we use the burning approach for the box covering problem \cite{song2007}, thus the boxes are defined for a central node or edge. Instead of calculating the distance between every pair of nodes in a box, the maximal distance to the central node or edge $r_B$ is given. This distance can then be related to the size of the box $r_{B} = (l_{B}-1)/2$ for a central node and $r_{B} = l_{B}/2$ for a central edge. The maximal chemical distance within a box of a given size $r_{B}$ is $2r_B$ for a central node and $2r_B - 1$ for a central edge. Although this problem can be simply stated, its solution is known to be NP-hard \cite{Garey1979}. It can be also mapped to a graph coloring problem in computer science \cite{Jensen1995} and has important applications, e.g., the calculation of fractal dimensions of complex networks \cite{Yook2005,Palla2005,Zhao2005,Goh2006,Moreira2006,song2006b} or the identification of the most influential spreaders in networks \cite{Kitsak2010}. Here we introduce an efficient algorithm for fractal networks which is capable to determine the minimum number of boxes for a given parameter $l_B$ or $r_B$. Moreover, we compare it for two benchmark networks with a standard algorithm used to approximately obtain the minimal number of boxes.
In principle, the optimal solution should be identified by testing exhaustively all possible solutions. Nevertheless, for practical purposes, this approach is unfeasible, since the solution space with its $2^N$ solutions is too large. Present algorithms like maximum-excluded-mass-burning \cite{song2007} and merging algorithms \cite{Locci2010} are based on the sequential addition of the box with the highest score, e.g., the score is proportional to the number of covered nodes, and the boxes with the highest score are sequentially included. Other algorithms are based on simulated annealing \cite{Zhou2007}, but without the guarantee of finding the optimal solution. {Even greedy algorithms end up with a similar number of boxes as the algorithms mentioned before \cite{Cormen2001}. The greedy algorithm sequentially includes a node to a present box, if all other nodes in this box are within the chemical distance $l_B$ and if there is no such box, a new box with the new node is created.} It is therefore believed that the results are close to the optimal result, although the real optimal solution is unknown.\\
%The greedy algorithm sequentially includes the box which covers the largest number of nodes into the solution and removes these covered nodes from all other boxes. It is therefore believed that the results are close to the optimal result, although the real optimal solution is unknown.\\
% First, we introduce the algorithm and then explain the main difference between the present state of the art algorithm and our optimal algorithm for a given distance $r_B$ before showing the possible improvement.
This paper is organized as follows. In Section II, we introduce the algorithm and then explain the main difference between the present state of the art algorithm and our optimal algorithm for a given distance $r_B$. In Section III, results for two benchmark networks are presented and the improvement in performance of our algorithm is quantitatively shown. Finally, in Section IV, we present conclusions and perspectives for future work.

\section{The Algorithm}
\begin{figure}
 \includegraphics[width=2.0cm,angle = 0]{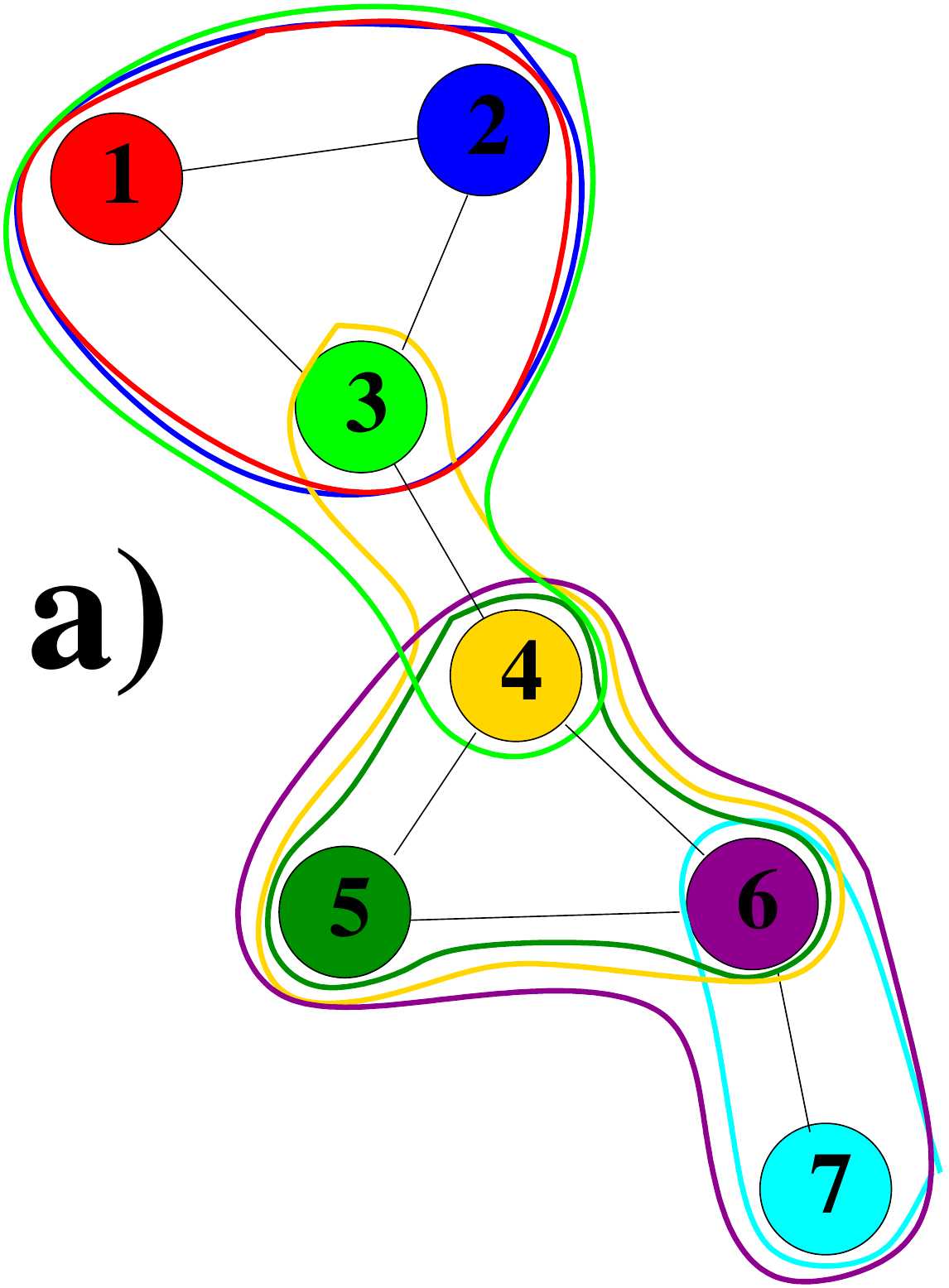}
 \includegraphics[width=2.0cm,angle = 0]{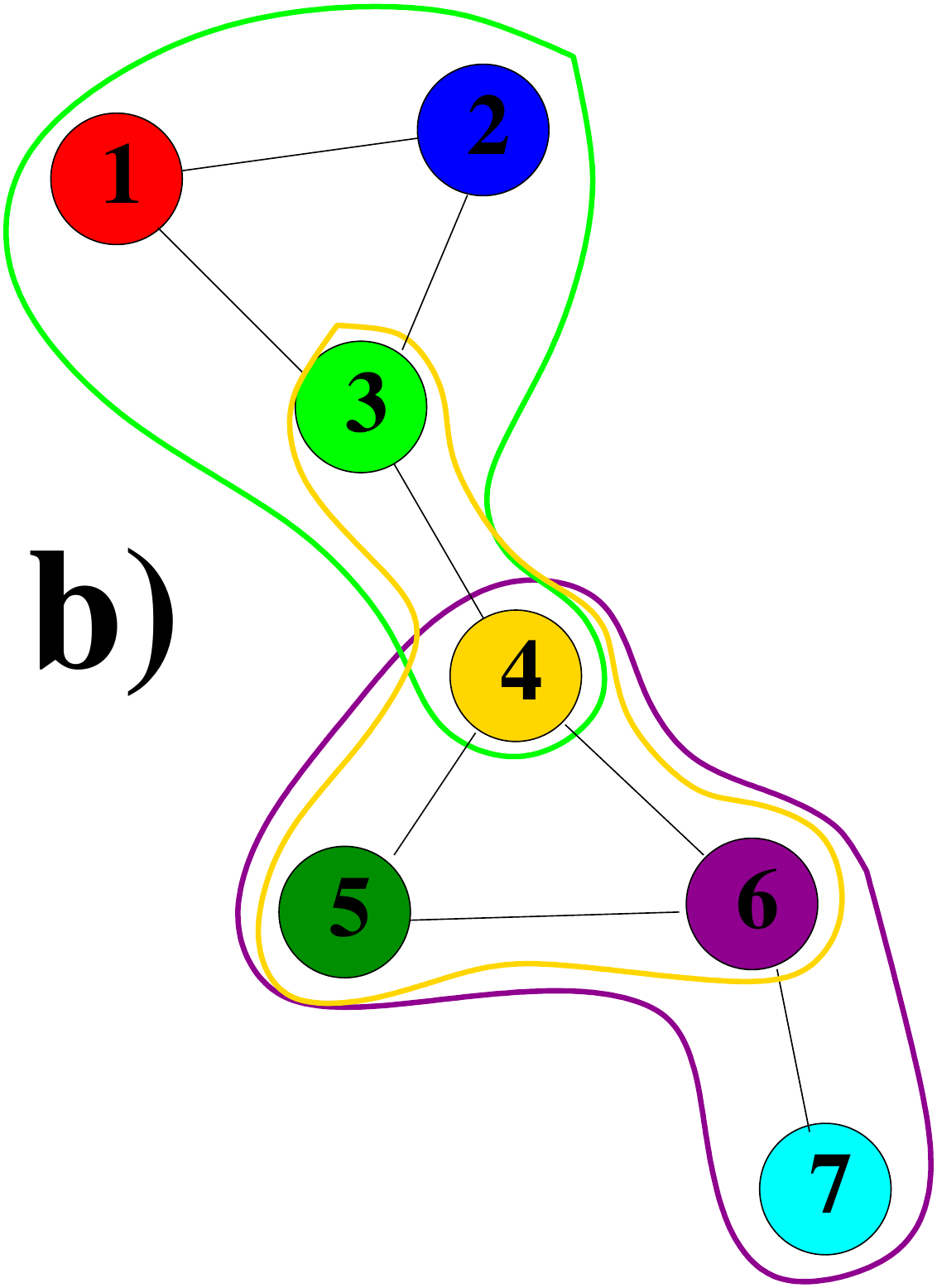}
 \includegraphics[width=2.0cm,angle = 0]{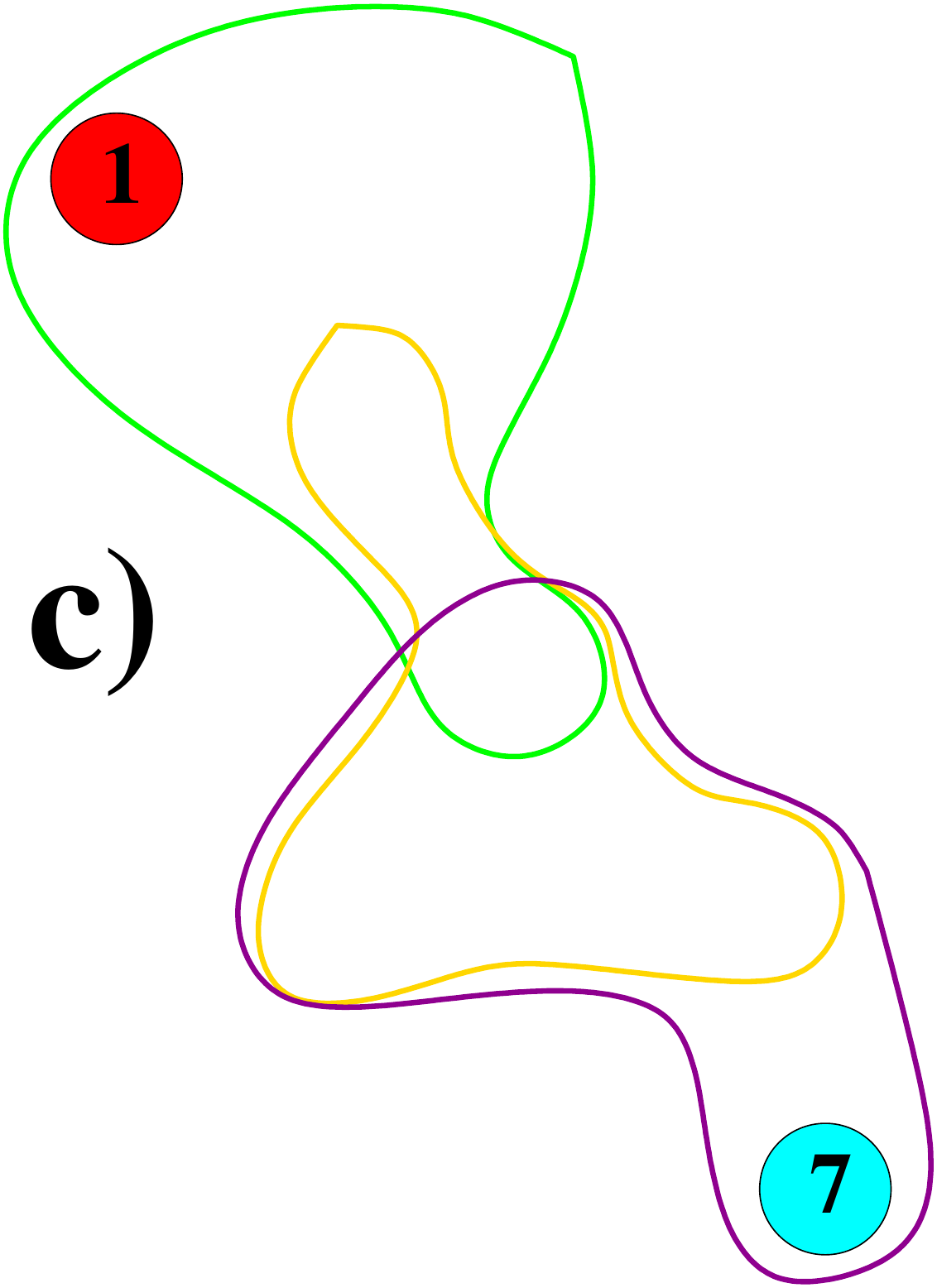}
 \includegraphics[width=2.0cm,angle = 0]{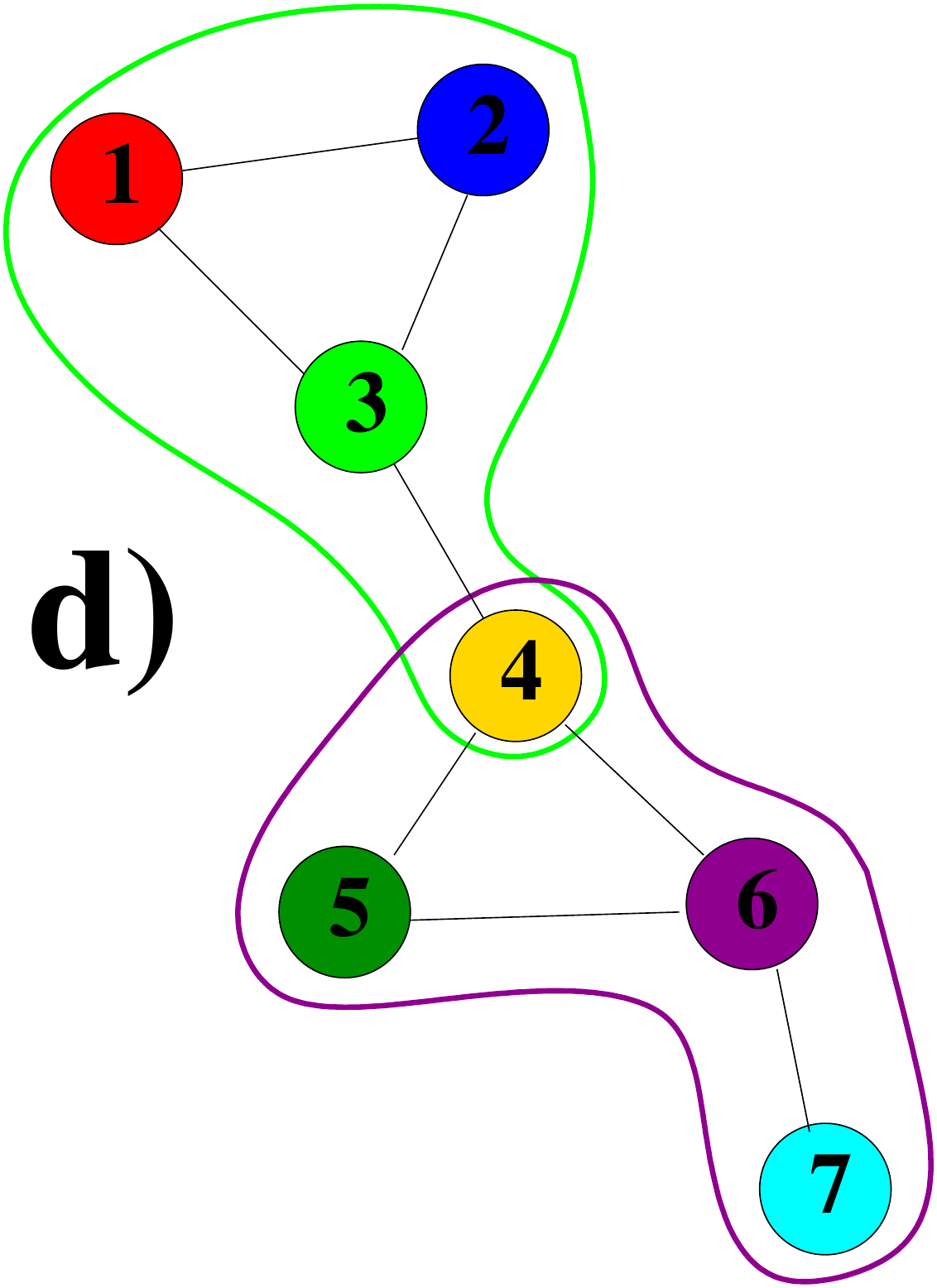}\\
 \includegraphics[width=2.0cm,angle = 0]{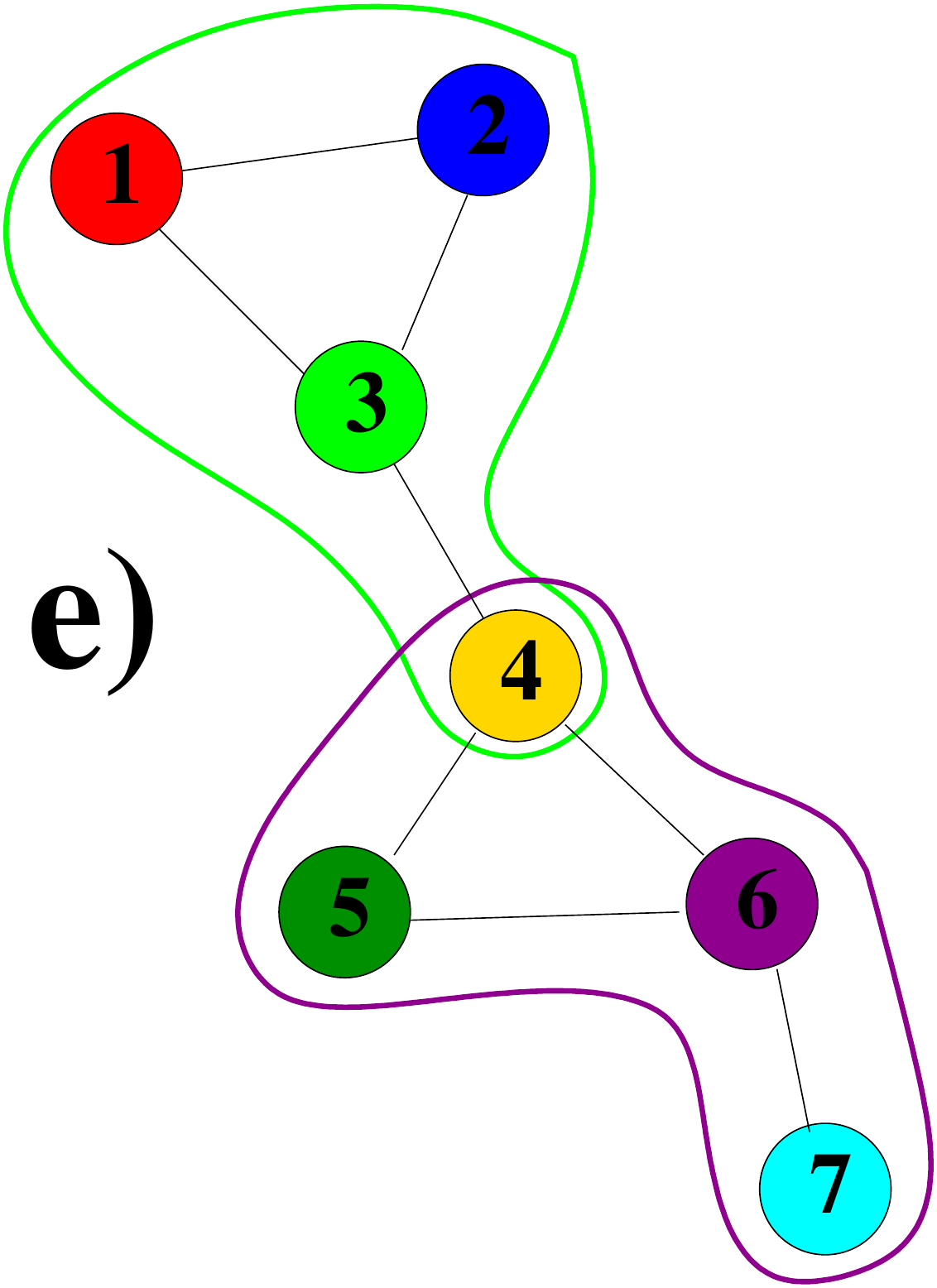}
 \includegraphics[width=2.0cm,angle = 0]{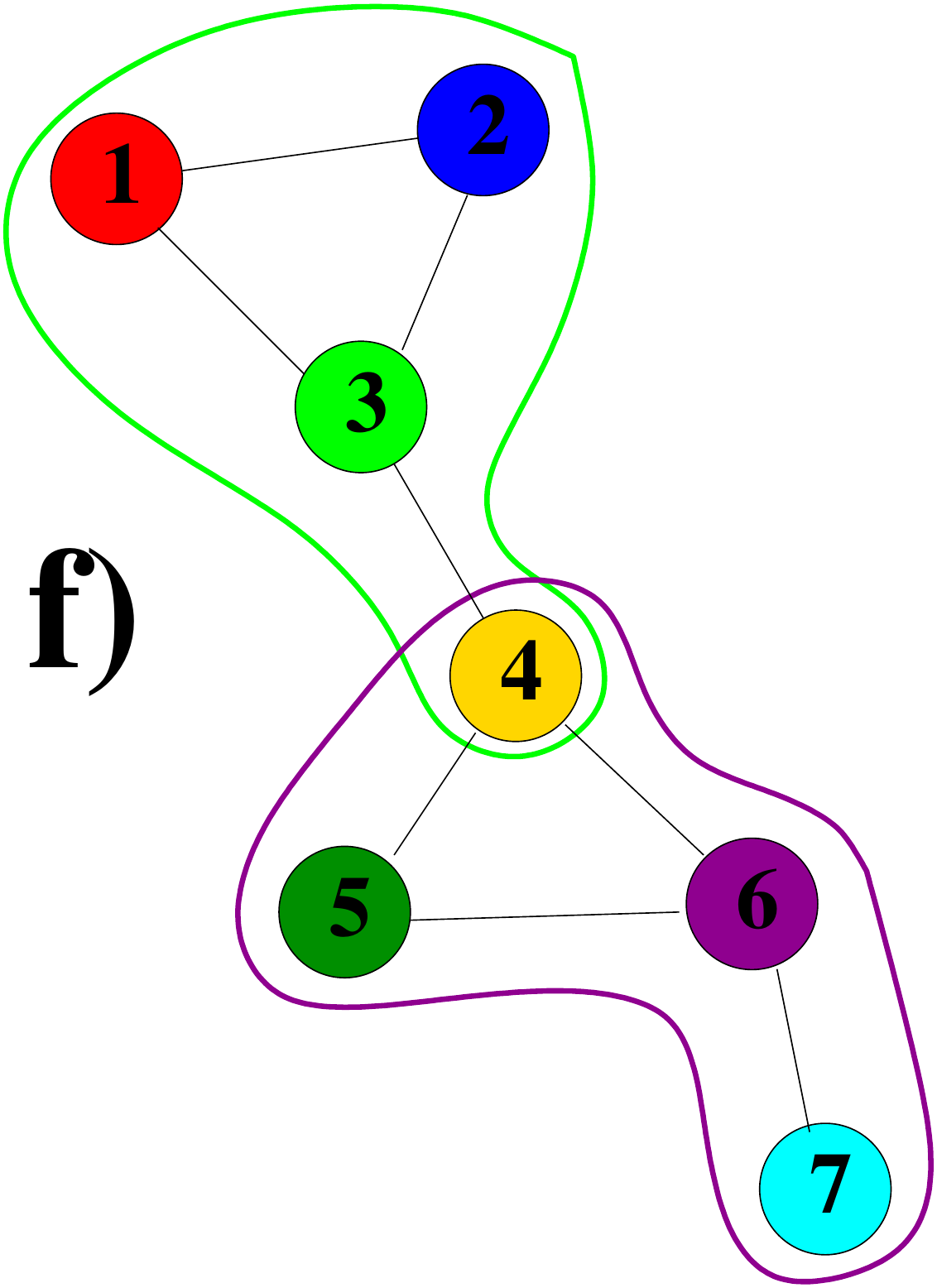}
 \includegraphics[width=2.0cm,angle = 0]{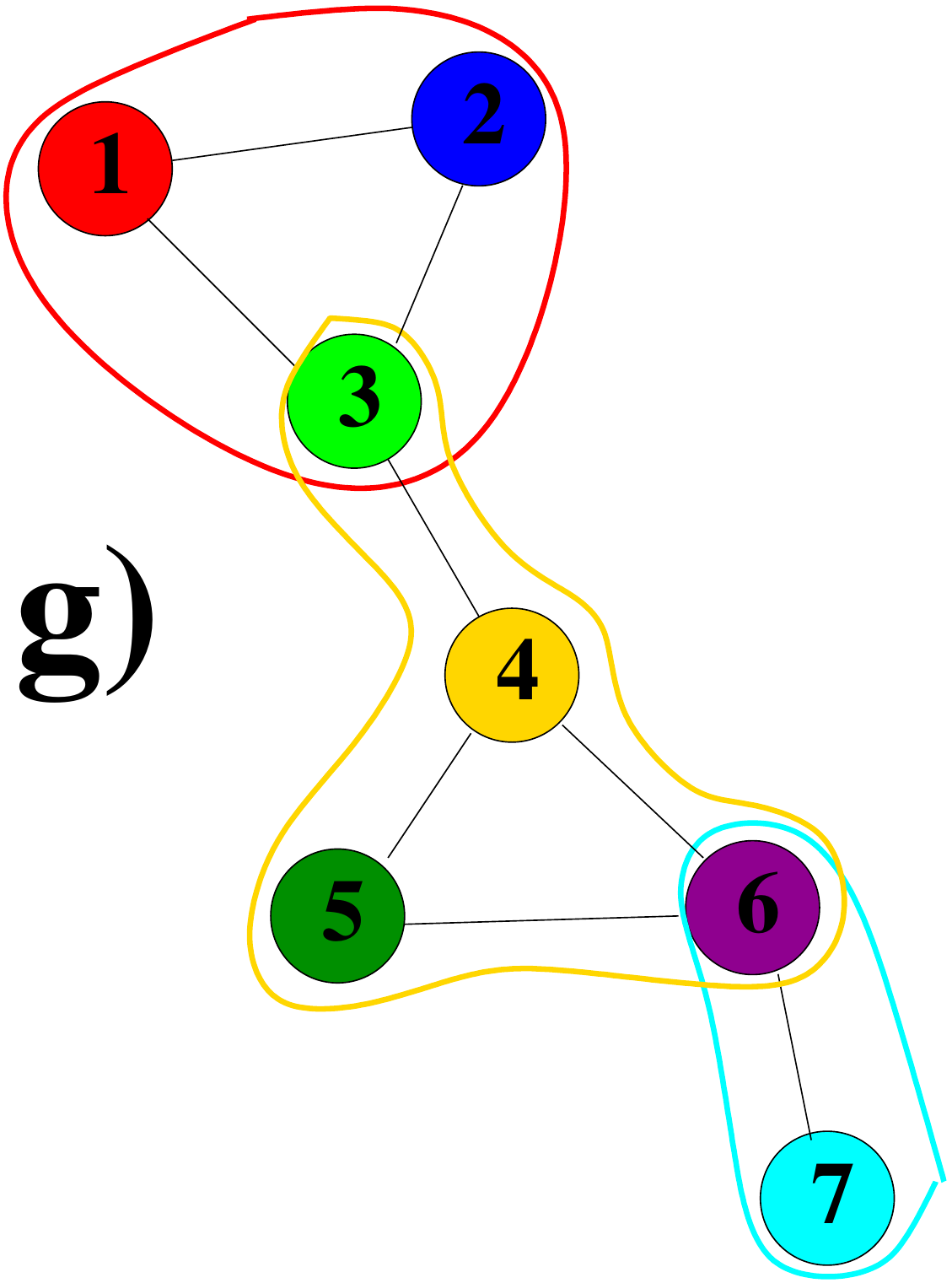}
 \caption{The box covering algorithm on a small example network for the box size $l_B = 3$ ($r_B = 1$ with a central node). Upper panel: a) Step 1: Calculation of all possible boxes. The color of the boxes corresponds to the node in its center. b) Step 2: All boxes that are fully contained in another box are removed. In this example the boxes $B_1$, $B_2$, $B_5$, and $B_7$ are removed. c) Step 3: All nodes which are in all boxes of another node are removed. In this example, nodes 2,3,4 are in the same box with node 1 as well as nodes 4,5,6 are in the same box with node 7. d) The final, optimal solution is shown on the right side.\\ Lower panel: The three possible solutions for the greedy box covering algorithm, based on the largest box sizes. In this case, the boxes are included to the solution according to the number of new covered nodes. Since three boxes $B_3$, $B_4$ and $B_6$ have the same number of nodes, the algorithm finds three different solutions e) ($B_3$,$B_6$), f) ($B_6$,$B_3$), and g) ($B_4$,$B_1$,$B_7$), where the last one is not optimal.}
\label{fig:1}
\end{figure}

\begin{figure}
 \includegraphics[width=2.5cm,angle = 0]{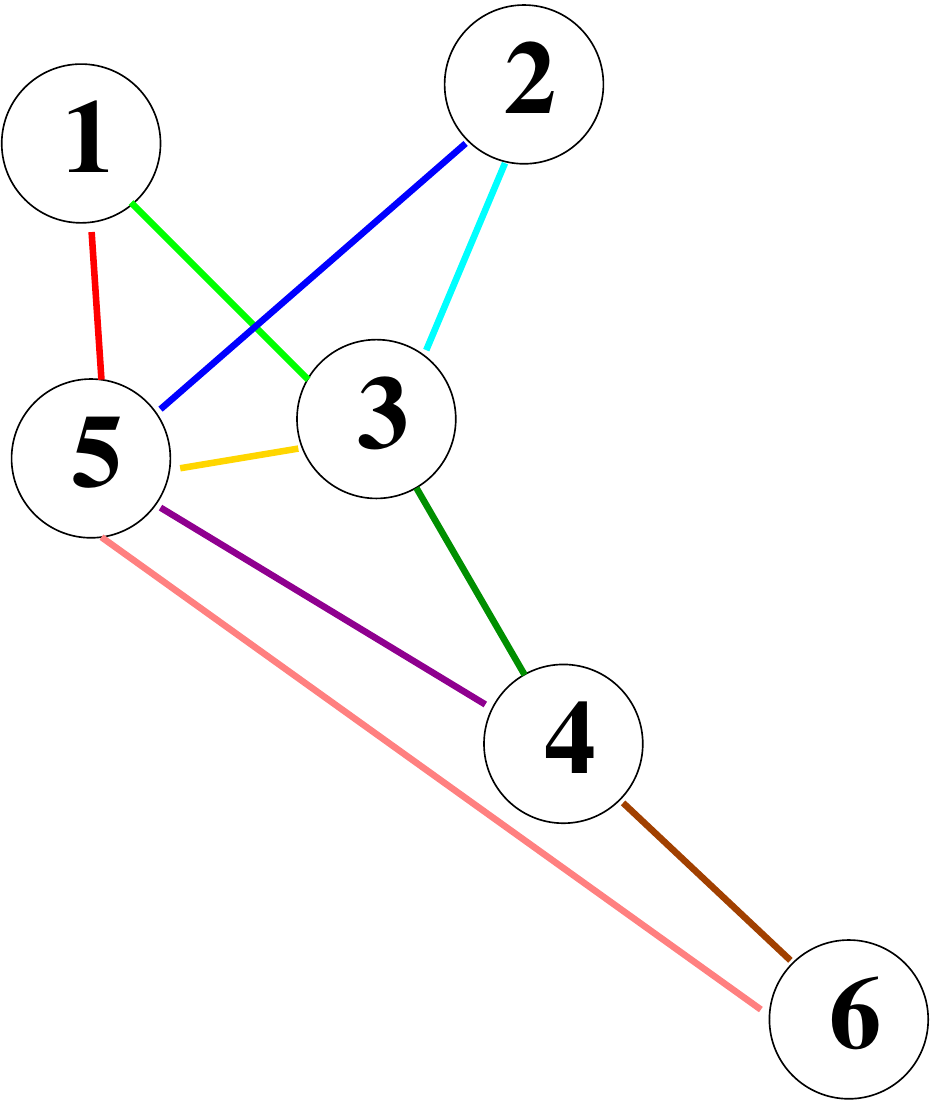}
 \includegraphics[width=2.5cm,angle = 0]{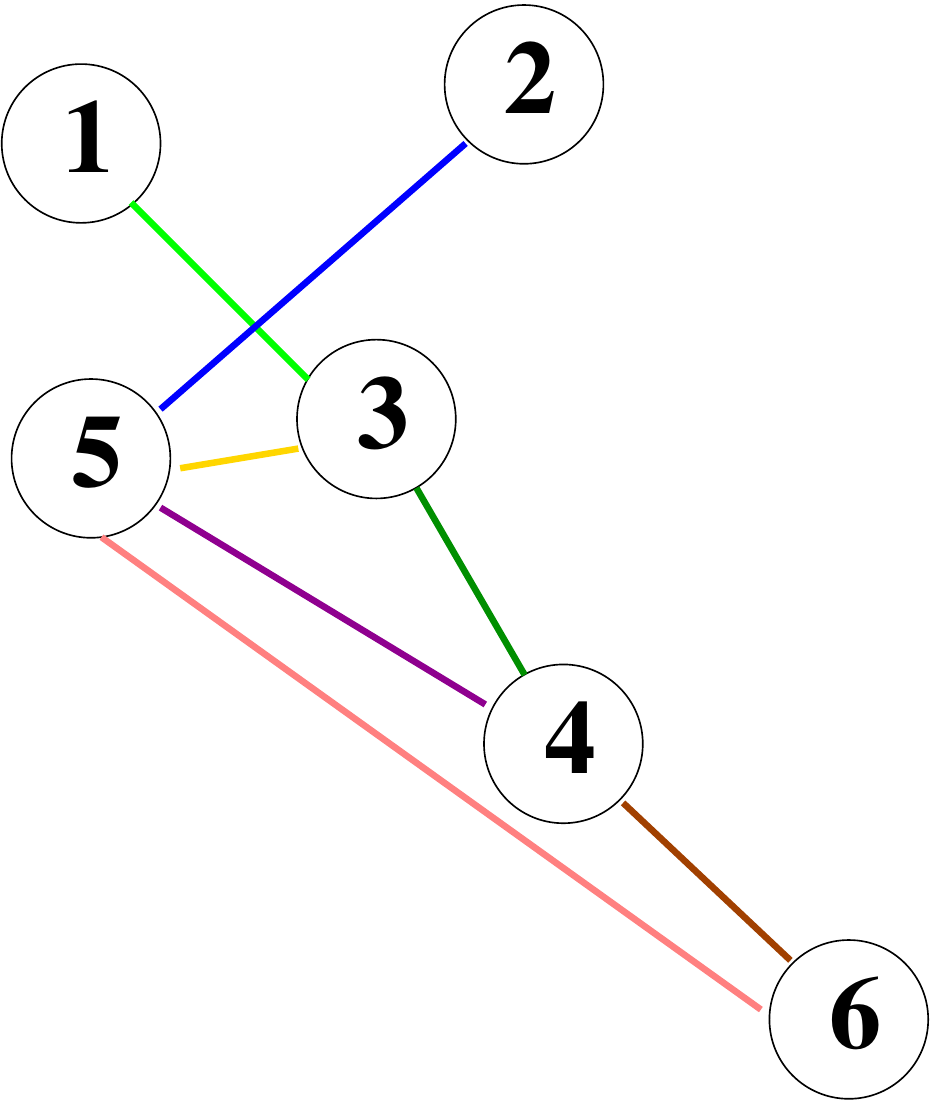}
 \caption{Step 4: In this example two nodes are in the same box, if they are connected with an edge. The two boxes between nodes 1 and 5 and between nodes 2 and 3 are removed according to rule 4.}
\label{fig:5}
\end{figure}

\begin{figure}
 \includegraphics[width=2.5cm,angle = 0]{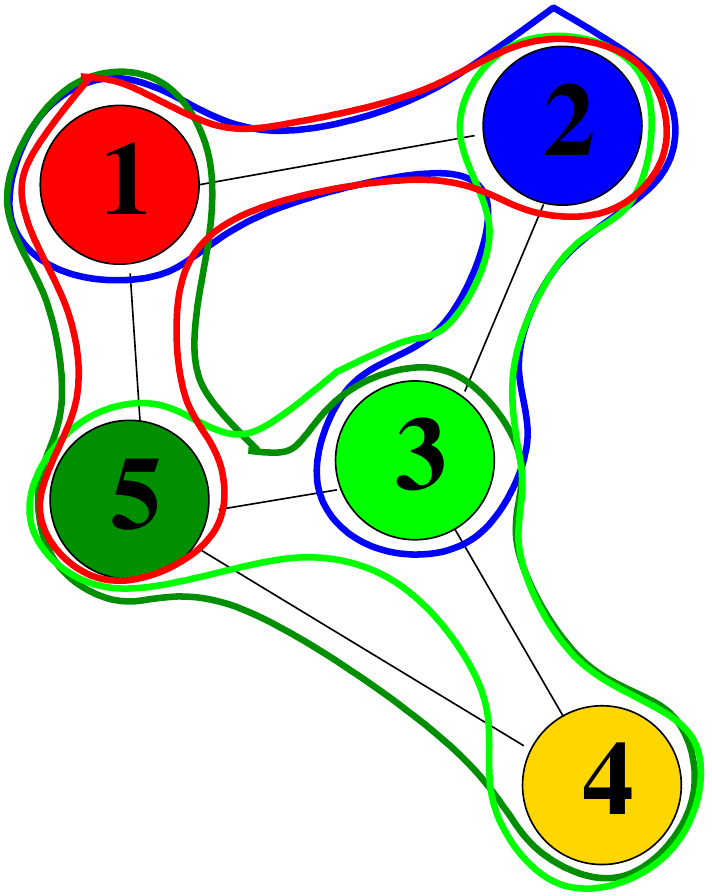}
 \includegraphics[width=2.5cm,angle = 0]{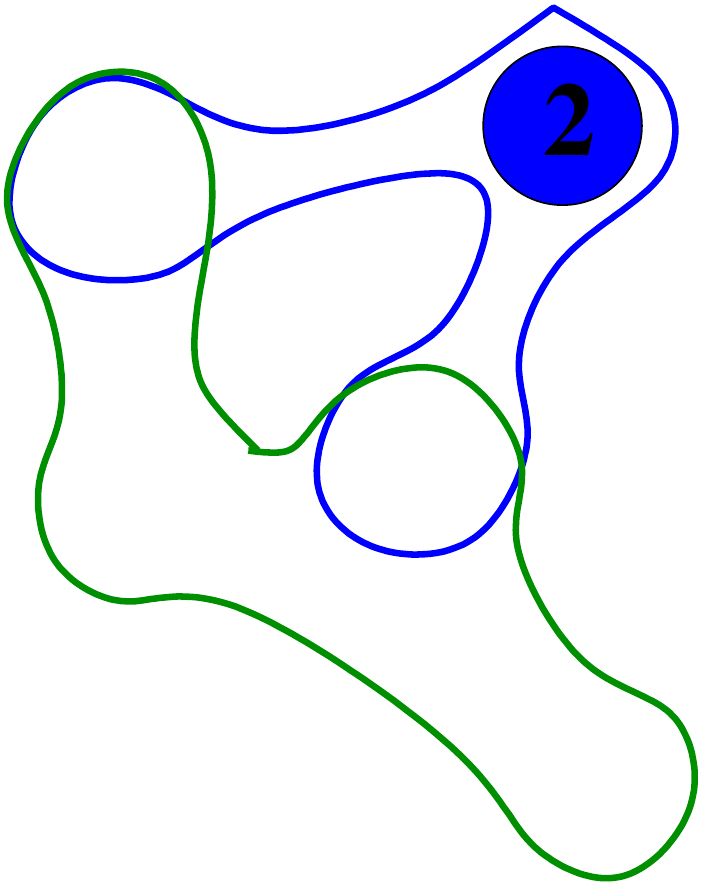}
 \includegraphics[width=2.5cm,angle = 0]{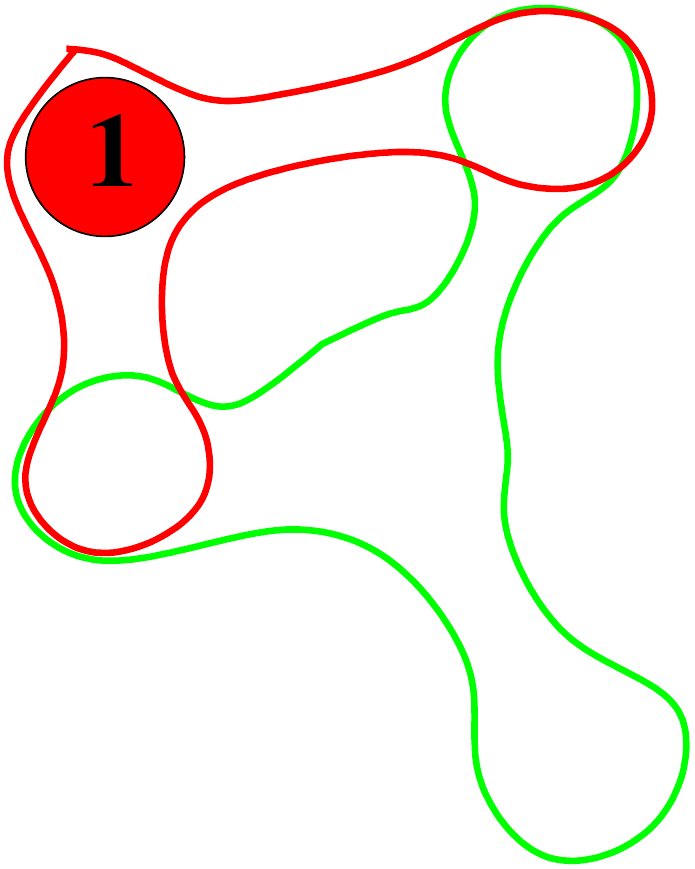}
 \caption{Step 8: Node 4 is covered by 2 circles (the minimal number of boxes) and the algorithm splits. The first sub-algorithm continues with box $B_5$ (middle), while the second one continues with box $B_3$(right).}
\label{fig:6}
\end{figure}

We use two slightly different algorithms for the calculation of the optimal box covering solution, one for odd values of $l_B$ and another for even values $l_B$. To get the results for an odd value, the following rules are applied:\\
\begin{enumerate}
\item Create all possible boxes: For every node $i$ create a box $B_i$ containing all nodes that are at most $r_B=(l_B-1)/2$ edges away. Node $i$ is called center of the box. An example is shown in Fig. \ref{fig:1}a.
\item Remove unnecessary boxes: Search and remove all boxes $B_i$ which are fully contained in another box $B_j$ (See Fig. \ref{fig:1}b).
\item Remove unnecessary nodes: For every node $i$, check all the boxes containing $i$: $B_{i_1}, ..., B_{i_n}$. If another node $j \neq i$ is contained in all of these boxes, remove it from \textit{all} boxes (see Fig. \ref{fig:1}c).
\item Remove pairs of unnecessary twin boxes: Find two nodes $i,j$ which are both in exactly two boxes of size two: $B_{i_1}=\{ i, k_1 \}$, $B_{i_2}=\{ i, k_2 \}$ and $B_{j_1}=\{ j, l_1 \}$, $B_{j_2}= \{ j, l_2 \}$. If $k_1=l_1$ and $k_2=l_2$, then $B_{i_2}$ and $B_{j_1}$ can be removed. If $k_1=l_2$ and $k_2=l_1$, then $B_{i_2}$ and $B_{j_2}$ can be removed. An example for this rule is shown in Fig. \ref{fig:5}. Note that such twin boxes also appear for $l_B > 2$ due to the removal of unnecessary nodes.
\item Search for boxes that must be contained in the solution: Add all boxes $B_i$ to the solution, which have a node $i$ only present in this box. Remove all nodes $j \neq i$ covered by $B_i$ from other boxes.
\item Iterate A: Repeat 2-5 until there is no node which is covered by a single box and is not part of the solution.
\item System split: Identify if the remaining network can be divided into subnetworks, such that all boxes in a subnetwork contain only nodes of this subnetwork. Then these subnetworks can be processed independent from each other.
\item System split: Find the node which is in the smallest number of boxes $N_{\text{boxes}}$, each of these boxes covers another set of nodes $B_i$. If there is more than one node fulfilling this criterion, chose the node which is covered by the largest boxes. Then the algorithm is divided into $N_{\text{boxes}}$ sub-algorithms, which can be independently calculated in parallel. By removing from each of the $N_{\text{boxes}}$ sub-algorithm another set of nodes $B_i$, all possible solutions are considered. An example for the splitting is shown in Fig. \ref{fig:6}. Since we want to identify only one optimal solution, we do not need to calculate the results of all sub-algorithms. As soon as one of the sub-algorithms identifies an optimal solution, we can skip the calculation of the others. Furthermore, the calculation of a sub-algorithm can be skipped, if the minimal number of required additional boxes reaches the number of the, so far, best solution of a parallel sub-algorithm.
\item Iterate B: Repeat 2-8 until no nodes are uncovered.
\item Identify the best solution: Chose the solution with the lowest number of boxes. This solution is optimal for a given $r_B$.
\end{enumerate}

To get the results for an even value of $l_B$ the first step is slightly different:
\begin{enumerate}
 \item Create all possible boxes: For every {\it edge} $i$ create a box $B_i$ containing all nodes that are at most $r_B=l_B/2$ {\it nodes} away. {\it Edge} $i$ is called center of the box.
\end{enumerate}
All other steps are the same as for the odd case. Note that the calculation for odd values scales with the number of nodes of the network $N$ and with the number of edges $M$ for even values.

\section{Results for two Benchmark Networks}
\begin{figure}
 \includegraphics[height=6.5cm,angle = 0]{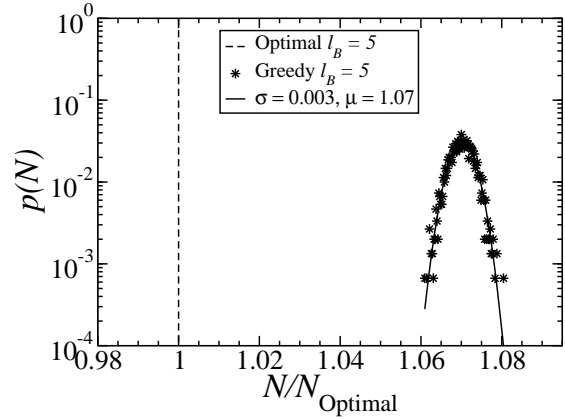}
 \caption{The distribution of minimal number of boxes $p(N)$ for the WWW network for $l_B = 5$ calculated through the greedy graph coloring algorithm for $1500$ different random node sequences. We have normalized the results by the optimal solution obtained from our algorithm. The distribution follows a normal distribution $p(x) \sim \exp(-(x-\mu)^2/(2\sigma^2))$ with $\mu = 1.07 \pm 0.01$ and $\sigma = 0.003 \pm 0.001$, thus approximately $10^{120}$ realizations are necessary to find the optimal solution with the greedy algorithm.}
 \label{fig:www1}
\end{figure}

Instead of sequentially including boxes, the idea of our algorithm is to remove all non-optimal boxes from the solution space ending up with a final, optimal solution. To reduce the huge solution space, our box covering algorithm uses two basic ingredients: 1) Unnecessary boxes from the solution space are discarded and the boxes which definitively belong to the solution are kept. 2) Unnecessary nodes from the network are discarded. These two steps reduce the solution space of a wide range of network types significantly, specially if they are applied in alternation as the removal of a box can lead to the removal of nodes and other boxes and vice-versa. Nevertheless these two steps do not necessarily lead to the optimal solution, thus the solution space has to be split into several possible sub-solution spaces. In each of these sub-solutions the first two steps are repeated. Note that the splitting does not reduce the number of possible solutions, thus only the first two steps reduce the solution space and in the worst case, the algorithm must calculate the entire solution space. In any case, for many complex networks iterating these three steps significantly reduces the solution space to a few solutions from which the optimal box covering can be obtained.\\
The remaining question is how to judge whether a box or node is necessary or unnecessary. On the one hand a box is unnecessary if all nodes of a box are also part of another box. This box can be removed, because the other box covers at least the same nodes and often additional nodes. On the other hand a box is necessary if a node is exclusively covered by this single box. This box has to belong to the solution, since only if the box is part of the solution, the node is covered.\\
In contrast, nodes can easily be identified as unnecessary. For example all nodes of a box, which is part of the solution, can be removed from all other boxes, since they are already covered. Additionally, if a node shares all boxes with another node, the other node can be removed, since the second node is always covered, if the first node is covered. These few rules are in principle sufficient to get the optimal solution, since our algorithm starts with {\it all} $2^N$ or $2^M$ (for central edges) possible solutions and discards unnecessary and includes necessary boxes.\\ %More details about our used algorithm are in the following section.\\
Although we only calculate results for undirected, unweighted networks, the algorithm can easily be extended to directed and weighted networks. In both cases only the initial step, the creation of boxes, is different. For directed networks, the box around a central node contains all nodes which are reachable with respect to the direction, while for weighted networks, the distance is the sum of the edge weights between the nodes.\\
\begin{figure}
 \includegraphics[height=6.5cm,angle = 0]{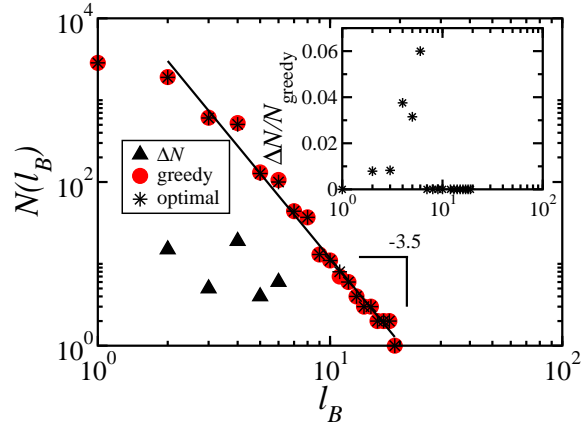}
 \caption{Comparison of the minimal number of boxes $N(l_B)$ for a given distance $l_B$ for the {\it E. Coli} network using the greedy graph coloring algorithm and our optimal algorithm. While the decay for both box covering methods is similar in the logarithmic plot, the minimal number of boxes is different. Although the difference $\Delta N = N_\text{greedy} - N_\text{optimal}$ seems to be small, the relative improvement $\Delta N / N_\text{greedy}$, which is shown in the inset, is significant for small distances $l_B < 7$. Note that the larger the box size the simpler the network can be covered with the optimal number of boxes. The straight line shows a power-law behavior, where the best fit for the fractal dimension is $d_B = 3.47 \pm 0.11$ for the greedy graph coloring and $d_B = 3.45 \pm 0.10$ for our optimal algorithm, respectively. Within the error bars both box-covering algorithms yield the same fractal dimension.}
 \label{fig:cellular}
\end{figure}
Next we show that our algorithm can also identify optimal solutions for large networks. Therefore, we have applied it to two different benchmark networks, namely the {\it E. Coli} network \cite{Makse}, with 2859 proteins and 6890 interactions between them, and the WWW network \cite{Albert1999}. We compare the results for the minimal box number $N(l_B)$ of our algorithm for different values of box sizes $l_B$ with the results of the greedy graph coloring algorithm \cite{}, as displayed in Fig. \ref{fig:cellular}. While the absolute improvement is rather small, the relative improvement is up to $6\%$ larger for $l_B < 7$. If the network is fractal, it should obey the relation, 
\begin{eqnarray}
N(l_B) \sim l_B^{-d_B},
\end{eqnarray}
where $d_B$ is the fractal dimension. Interestingly, it seems that the fractal dimension $d_B = 3.47 \pm 0.11$ from the greedy algorithm and $d_B = 3.45 \pm 0.10$ from our optimal algorithm of the network is nearly unaffected by the choice of the algorithm. Note that for $l_B = 11$, due to the fact that the boxes are calculated based on the definition of a central node or edge, we have one more box. The simplest case where such difference occurs is in a chain of four connecting nodes (1-2, 2-3, 3-4, 4-1). All nodes have the chemical distances of two to each other ($l_B = 3$), however it is not possible to draw a box around a node with radius one ($r_B = 1$), which contains all nodes.\\
\begin{figure}
 \includegraphics[height=6.5cm,angle = 0]{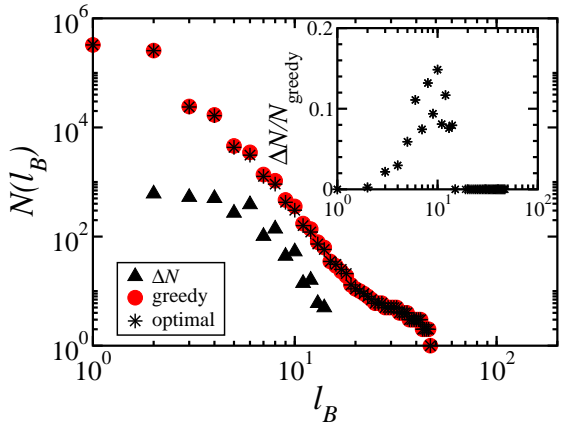}
 \caption{The minimal number of boxes $N(l_B)$ as a function of the distance $l_B$ for the WWW network calculated through the greedy graph coloring algorithm and our optimal algorithm. While the fractal dimension for both box covering methods is nearly similar, the minimal number of boxes is different. The difference $\Delta N = N_\text{greedy} - N_\text{optimal}$ as well as the relative improvement $\Delta N / N_\text{greedy}$, which is shown in the inset, are significant for $l_B < 16$. For this network a maximal relative improvement of about $15\%$ can be obtained.}
 \label{fig:www}
\end{figure}
The second example is the WWW network, containing 325729 nodes and 1090108 edges. As in the previous case, our algorithm outperforms the state of the art algorithm, but yields similar fractal behavior, as shown in Fig. \ref{fig:www}. For intermediate box sizes $l_B < 16$, we have a large improvement since up to $15\%$ and up to $611$ fewer boxes are needed. For $l_B = 16, 17, 18$ we have two box more, like in the {\it E. Coli} network case due to the two definitions of the box covering problem, while for larger $l_B$ both algorithm give similar results. Interestingly, it seems that the improvement for even distances $l_B$ (for central edges) is significantly larger than for odd distances $l_B$ (for central nodes).\\
{In Fig. \ref{fig:www1} we show the influence of the sequence of adding nodes to the boxes on the results of the greedy algorithm. While the results of Fig. \ref{fig:www} are the minimal values obtained from 50 independent starting sequences, we calculated 1500 realizations for a single box size $l_B = 5$. The difference between the improvement is with $N_\textrm{greedy}/N_\textrm{optimal} = 6.3\%$ and $N_\textrm{greedy}/N_\textrm{optimal} = 6.1\%$ rather small. The gap between the optimal solution and the greedy algorithm is too large, thus for practical purposes, the greedy algorithm will never find the optimal solution for this box size.}\\
The results for these two benchmark networks demonstrate that our algorithm is more effective than the state of the art algorithms. Nevertheless, due to the rapid decay of the number of boxes for larger box sizes, the fractal dimension of the two benchmark networks is only slightly different when using the optimal box-covering algorithm in comparison with other algorithms.\\
%the calculation but the application of the box covering on these networks for the calculation of the fractal dimension is rather ???(conceptional).\\

\section{Conclusions}
In closing, we have presented a box-covering algorithm, which outperforms the known previous ones. We have also compared our algorithm with the state of the art methods for different benchmark networks and detected substantial improvements. Moreover the obtained solutions are optimal as a result of the algorithm design, if the box size is defined as the maximal distance $r_B$ to the central node or edge. For example, our approach can be useful for designing optimal commercial distribution networks, where the shops are the nodes, the storage facilities the box centers and the radius is related to the boundary conditions, like transportation cost or time.
%In closing, we have presented a new box-covering algorithm, which outperforms the know ones and successfully used it to identify the key industries in development assistance. We have also compared our algorithm with state of the art ones on different benchmark networks to show the huge improvement potential. Moreover the obtained solutions are optimal, if the nodes of each box have a maximal distance of $r_B$ to the center of the box. 

\section{Acknowledgment}
We acknowledge financial support from the ETH Competence Center 'Coping with Crises in Complex Socio-Economic Systems' (CCSS) through ETH Research Grant CH1-01-08-2 and by the Swiss National Science Foundation under contract 200021 126853. We also thank the Brazilian agencies CNPq, CAPES, FUNCAP and the INST-SC for financial support.

\end{document}